\documentclass[conference]{IEEEtran}
\IEEEoverridecommandlockouts
\usepackage{amssymb}
\usepackage{algorithmic}
\usepackage{textcomp}
\usepackage{xcolor}
\def\BibTeX{{\rm B\kern-.05em{\sc i\kern-.025em b}\kern-.08em
    T\kern-.1667em\lower.7ex\hbox{E}\kern-.125emX}}

\usepackage{amsmath,graphicx,amsfonts,comment,array,multirow, cite, here}
\usepackage{subfig}

\usepackage{xcolor,url}

\def\ERB{\mathrm{ERB}}
\def\Min{M^{\mathrm{(in)}}}
\def\Mout{M^{\mathrm{(out)}}}
\def\mpgtf{\mathrm{(MP\text{-}GTF)}}
\def\thline{\noalign{\hrule height 1.0pt}}
\def\tthline{\noalign{\hrule height 1.4pt}}
\def\target{\mathrm{(target)}}

\def\mysection#1{\section{#1}}
\def\mysubsection#1{\subsection{#1}}

\begin{document}
\title{Sampling-Frequency-Independent Audio Source Separation Using Convolution Layer Based on Impulse Invariant Method
\thanks{This work was supported by JSPS KAKENHI Grant Number JP20K19818.}
}
%
\author{Koichi Saito$^\dagger$, Tomohiko Nakamura$^\dagger$, Kohei Yatabe$^\ddagger$, Yuma Koizumi$^\star$, Hiroshi Saruwatari$^\dagger$
\\
$^\dagger$Graduate School of Information Science and Technology, The University of Tokyo, Tokyo, Japan \\
$^\ddagger$Department of Intermedia Art and Science, Waseda University, Tokyo, Japan \\
$^\star$NTT Corporation, Tokyo, Japan 
}

\maketitle
\begin{abstract}
Audio source separation is often used as preprocessing of various applications, and one of its ultimate goals is to construct a single versatile model capable of dealing with the varieties of audio signals.
Since sampling frequency, one of the audio signal varieties, is usually application specific, the preceding audio source separation model should be able to deal with audio signals of all sampling frequencies specified in the target applications.
However, conventional models based on deep neural networks (DNNs) are trained only at the sampling frequency specified by the training data, and there are no guarantees that they work with unseen sampling frequencies.
In this paper, we propose a convolution layer capable of handling arbitrary sampling frequencies by a single DNN. 
Through music source separation experiments, we show that the introduction of the proposed layer enables a conventional audio source separation model to consistently work with even unseen sampling frequencies.
\end{abstract}
\vspace{-0.2\baselineskip}
\mysection{Introduction}
\label{sec:intro}
\vspace{-0.2\baselineskip}

\begin{figure*}[ht]
\centering
\includegraphics[width=1.9\columnwidth]{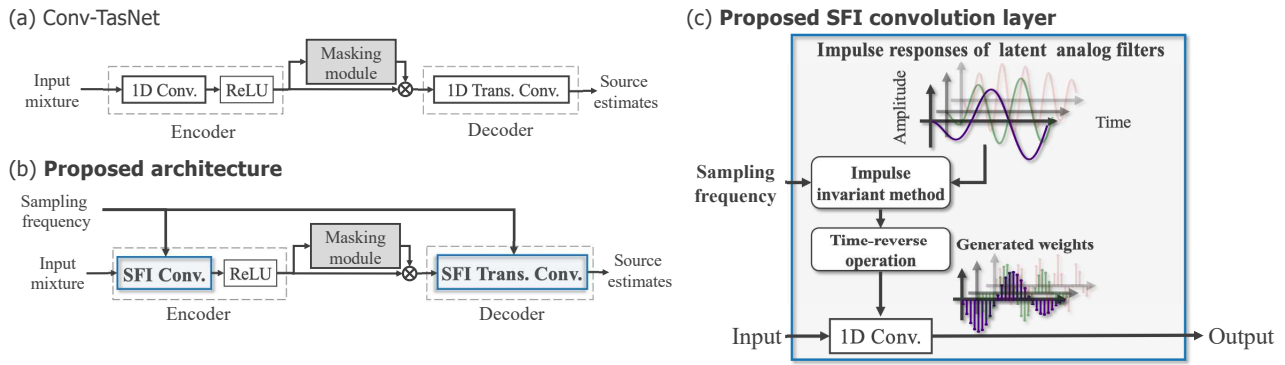} 
\caption{Architectures of (a) Conv-TasNet \cite{convtas} and (b) proposed model, and (c) illustration of proposed SFI convolution layer.}
\label{fig:sfc}
\end{figure*}

Audio source separation is a technique for extracting individual sources from a mixture signal.
It is one of the fundamental techniques for various audio applications including music remixing, automatic music transcription, and automatic speech recognition.
The recent development of source separation has been built upon machine leaning techniques using the deep neural network (DNN) \cite{8462507, 7952154, convtas, waveunet, defossez2019music, 9053934,mpgtf,liu2020channelwise,Takeuchi2019ICASSP}.
Since source separation is often utilized as a preprocess of another task, one of the ultimate developmental goals is to construct a single universal DNN that can be utilized as the preprocessor for any application.
To realize such an almighty source separator, every variety of applications and conditions must be handled by a single DNN.

One important but often unnoticeable variety of audio signals is sampling frequency.
It is usually application specific, and hence a preprocessor must be designed for the sampling frequency specified by the following application.
For example, for music remixing and editing, $44.1$ and $48$ kHz are usually used as sampling frequencies to cover the entire human audible range \cite{defossez2019music, liu2020channelwise}. 
This is because these applications are aimed at human listeners.
In contrast, the applications aimed at the recognition of the contents contained in audio signals do not require such full-band information.
For example, beat tracking may use $16$ kHz \cite{Krebs2016DownbeatTU}, automatic music transcription may use $11.025$ and $22.05$ kHz \cite{7416164, 9052987}, and automatic speech recognition may use $8$ and $16$ kHz \cite{Yu2013ICLR,8639610,8574905}.
A versatile preprocessor must be able to handle signals sampled at all of these sampling frequencies.

However, ordinary DNNs cannot handle audio signals sampled at various sampling frequencies.
Conventional DNN-based models work well for the sampling frequency specified by the training data \cite{8462507, 7952154, convtas, waveunet, defossez2019music, 9053934,mpgtf,liu2020channelwise,Takeuchi2019ICASSP}.
The parameters of a DNN are trained to adapt for the training dataset, and thus there is no guarantee of applicability for signals that are sampled at the other (unseen) sampling frequencies.
This is because the layers utilized in a DNN are not designed for multiple sampling frequencies.
In fact, the sampling frequency has not been considered as a parameter of a DNN, but it is implicitly given by the training dataset.
In order to realize a DNN that consistently works for any sampling frequency, a DNN must be designed as a \textit{sampling-frequency-independent (SFI)} network. 

In this paper, we propose an SFI convolution layer for the handling of arbitrary sampling frequencies by a single DNN.
The key idea behind the proposed layer is to consider the connection between a digital filter and a convolution layer.
From a signal processing viewpoint, we can interpret a convolution layer as a collection of time-reversed digital finite impulse response (FIR) filters.
Therefore, a filter design technique can be utilized to design a convolution layer.
In this paper, we consider the \textit{impulse invariant method} (see Chap.~7 in \cite{Oppenheim2001DSP}), in which a digital filter is designed by sampling an analog filter.
On the basis of this analog-to-digital filter conversion, we introduce latent analog filters into a convolution layer.
Since an analog filter is independent of sampling frequency, we can construct an SFI convolution layer via the analog representation of a filter, where the impulse invariant method determines its sampling frequency afterward.
The proposed SFI layer can be trained by parametrizing the analog filter as a differentiable function.
By incorporating the proposed layer into one of the state-of-the-art source separation models, we also propose an SFI audio source separation model.

\vspace{-0.2\baselineskip}

\mysection{Conventional Models}
\label{sec:previousworks}
\vspace{-0.15\baselineskip}

\mysubsection{Conv-TasNet \cite{convtas}}
\label{ssec:convtasent}

Conv-TasNet is a recent time-domain DNN for audio source separation that works well for speech \cite{convtas} and music source separation \cite{metatas,defossez2019music}.
Since the architectures of Conv-TasNet are slightly different in those three papers, we adopt the Conv-TasNet architecture for music source separation defined in \cite{metatas}, as illustrated in Fig.~\ref{fig:sfc}(a).
Conv-TasNet consists of a pair of an encoder and a decoder and $C$ source-specific masking modules, where $C$ denotes the number of sources.
The encoder and decoder imitate a traditional time-frequency transform (e.g., the short-time Fourier transform) and its inverse transform.
The encoder transforms a monaural time-domain signal into an $N$-channel latent representation by a one-dimensional (1D) convolution layer (with kernel size $L$ and stride $W$) and the rectified linear unit (ReLU).
Each masking module estimates a mask for the target source from the latent representation.
It comprises $R$ convolution blocks, each of which consists of $X$ 1D dilated convolution layers with an exponentially increasing dilation factor.
The details of the convolution block are shown in \cite{convtas}.
The decoder converts the masked latent representations into the separated time-domain signals by a 1D transposed convolution layer with kernel size $L$ and stride $W$.

\vspace{-0.15\baselineskip}
\mysubsection{Multi-phase Gammatone Filter \cite{mpgtf}}
\label{ssec:multiphasegammatonefilter}
In \cite{mpgtf}, the multi-phase gammatone filter (MP-GTF) was introduced to design the weights of the convolution layer of the encoder of Conv-TasNet, which improved the speech separation performance.
The impulse response of the MP-GTF is given by
\begin{equation}
g^{\mpgtf}(t) = a\,t^{p-1}e^{-2\pi bt}\cos{(2\pi ft+\phi)}, \label{eq:agtf}
\end{equation}
where $a$ denotes the amplitude, $p$ the filter order, $b$ the bandwidth, $f$ the center frequency, and $\phi$ the phase shift.
The parameter $b$ is given by $b=\ERB(f)/1.57$, where $\ERB(f)= 24.7 + f/9.265$.
By sampling $L$ points from $g^{\mpgtf}(t)$ for various $f$ and $\phi$, we obtain $N$ discrete-time impulse responses of length $L$ and concatenate them along the channel axis to form the weights as a $1\times N\times L$ tensor.
The convolution layer of the encoder is followed by the ReLU nonlinearity, which blocks the negative values of the MP-GTF output and discards the information of the input signal.
To avoid a lack of information, $g^{\mpgtf}(t)$ is used together with its phase-reversed version, i.e., with the phase shift $\phi+\pi$ \cite{mpgtf}.
\vspace{-0.15\baselineskip}
\mysubsection{Multiple-sampling-frequency Training \cite{Yu2013ICLR,8639610,8574905,metatas}}

There exist a few methods of training DNNs using audio signals sampled at multiple sampling frequencies \cite{Yu2013ICLR,8639610,8574905,metatas}.
In \cite{Yu2013ICLR,8639610}, an automatic speech recognition (ASR) model was trained using audio signals sampled at $8$ and $16$ kHz, where the part of input features corresponding to the missing frequency band was padded by zeros.
In \cite{8574905}, to compensate the missing frequency band, an ASR model was jointly trained with a bandwidth expansion model.
The music source separation model presented in \cite{metatas} was constructed by stacking three Conv-TasNets that account for sampling frequencies of $8$, $16$, and $32$ kHz.
The Conv-TasNets for $16$ and $32$ kHz estimate the source signals of the target sampling frequencies, referring to the masked latent representations obtained with the lower sampling frequencies.

While these training methods are valid for the trained sampling frequencies, they are not guaranteed to work with unseen sampling frequencies.
In contrast, we explicitly define an SFI structure to handle any sampling frequency without retraining as shown later in Section~\ref{sec:proposedmethod}.

\vspace{-0.1\baselineskip}
\mysection{Proposed Model}
\label{sec:proposedmethod}
\vspace{-0.1\baselineskip}

\mysubsection{Sampling-frequency-independent (SFI) Convolution Layer}
\label{ssec:generativeprocess}
To realize an SFI network, we introduce latent analog filters and analog-to-digital filter conversion into a convolution layer.
By interpreting the weights of a convolution layer as a collection of time-reversed digital FIR filters, we consider them from a signal processing viewpoint.
The digital filters are inherently sampling frequency dependent, whereas analog filters are SFI owing to their definition in the continuous time domain.
Focusing on this fact, we introduce latent analog filters behind a convolution layer so that its weights can be adjusted using the sampling frequency of an input signal.

As shown in Fig.~\ref{fig:sfc}(c), the proposed layer consists of the usual 1D convolution layer and impulse responses of $\Min \Mout$ analog filters defined in the continuous time domain, where $\Min$ and $\Mout$ are the input and output channel sizes, respectively.
The generating process of the proposed layer consists of three steps.
Given the sampling frequency of an input signal, the proposed layer (i) generates a discrete-time impulse response of length $L$ from each analog filter, (ii) stacks the time-reversed versions of these discrete-time impulse responses to form the weights as an $\Min \times \Mout \times L$ tensor, and (iii) works as the usual convolution layer using them.
Since steps (i) and (ii) depend only on the sampling frequency and the continuous-time impulse responses, we only need to perform them once (before the features are input) whenever the sampling frequency changes.

For step (i), we employ the impulse invariant method to generate digital FIR filters from their analog counterparts.
Note that while this method is originally for the design of infinite impulse response filters, we can use it for the digital FIR filter design.
Let us denote the sampling period as $T$, a discrete time index as $l=1,\cdots,L$, and the continuous time as $t\in\mathbb{R}$.
The impulse invariant method generates a discrete-time impulse response $h[l]$ from an analog filter $g(t)$ so that the sampled instants coincide:
\begin{equation}
h[l] = Tg(lT).
\label{eq:invimp}
\end{equation}
Changing $T$ yields an impulse response for different sampling frequencies $1/T$.
By stacking the generated impulse responses, the weights for the convolution layer is obtained in step (ii).
Similarly, an SFI version of a transposed convolution layer (SFI transposed convolution layer) is given by changing the convolution layer in the SFI convolution layer to the transposed convolution layer.

For the analog filter $g(t)$, we can use the MP-GTF in Eq.~\eqref{eq:agtf}.
The continuous-time impulse responses can be different for each channel, and hence, hereafter, a channel subscript $m$ is added to $g^{\mpgtf}(t)$ and its parameters: $g^{\mpgtf}_{m}(t), a_{m}, p_{m}, b_{m}, f_{m}$, and $\phi_{m}$.
Whereas all parameters of $g^{\mpgtf}_{m}(t)$ were fixed in \cite{mpgtf}, we propose to train $f_m$ and $\phi_m$ jointly with the other DNN components by the commonly-used backpropagation algorithm. 

The gradient of $h[l]$ can be computed in the same manner as the usual convolution layer. 
Since the gradient of $h[l]$ equals that of $g(lT)$ owing to Eq.~\eqref{eq:invimp} and $g^{\mpgtf}_m(t)$ is differentiable with $f_m$ and $\phi_m$, the gradients of the trainable parameters of $g^{\mpgtf}_m(t)$ can be computed by the chain rule.
These computations can be easily implemented by defining the forward computation process of the proposed layer owing to the automatic differentiation mechanism installed in modern deep learning frameworks (e.g., PyTorch and TensorFlow).

\vspace{-0.15\baselineskip}
\mysubsection{Aliasing Reduction Technique}
\label{sssec:heuristictechniqueforaliasing}
Since the impulse invariant method simply performs sampling to an analog filter, aliasing occurs in the obtained digital filters.
As reported in \cite{Zeiler2014ECCV,Gong2018Interspeech,9053934}, aliasing causes degradation of the DNN performance, and thus we introduce an aliasing reduction technique. 
Since the energy of aliased components is distributed above the Nyquist frequency, we propose to set the weights of the $m$th channel to zero whenever the center frequency $f_m$ of the corresponding analog filter is above the Nyquist frequency.
This aliasing reduction technique is important for the proposed layer when it is utilized with low sampling frequencies, as shown later in Section~\ref{sec:experimentalevaluation}. 
\vspace{-0.15\baselineskip}
\mysubsection{Application of Proposed Layers to Conv-TasNet}
\label{ssec:adaptingsfv}

As shown in Fig.~\ref{fig:sfc}(b), we build an SFI audio source separation model by incorporating the proposed layers into Conv-TasNet \cite{convtas}.
The convolution layer of the encoder and the transposed convolution layer of the decoder are respectively replaced with the SFI convolution and transposed convolution layers.
The masking modules are the same as in \cite{convtas}.

For our model, we should modify the kernel size $L$ and stride $W$ in accordance with the sampling frequency during inference.
As described in Section~\ref{ssec:convtasent}, the encoder and decoder can be interpreted as the time-frequency transform and its inverse transform.
With this interpretation, $L$ and $W$ correspond to the frame length and the frame shift, respectively.
Hence, as the sampling frequency doubles, $L$ and $W$ should double to make the representation consistent for the masking module.
For this reason, we determine $L$ and $W$ for each target sampling frequency to keep the frame length and shift unchanged in the continuous time domain.
By replacing all convolution layers in the masking modules with the SFI convolution layers, this issue might be resolved.
However, we left it as a future work because additional care regarding the combination with other layers (e.g., group normalization \cite{Wu2018ECCV}) must be considered.

\vspace{-0.2\baselineskip}

\mysection{Experimental Evaluation}
\vspace{-0.1\baselineskip}
\label{sec:experimentalevaluation}
\begin{table}[t]
\centering
\caption{Features of proposed methods and Conv-TasNet}
{
\begin{tabular}{c|ccc} \tthline
\multirow{2}{*}{Method} & \multirow{2}{*}{$g_m(t)$} & Samp. freq. & Aliasing \\ 
& & adapt. & reduction \\ \thline
Conv-Tasnet \cite{convtas} & - & No & - \\
T-MP-GTF & $g_m^{\mpgtf}(t)$ & No & No \\
Proposed & $g_m^{\mpgtf}(t)$ & \textbf{Yes} & No \\
Proposed+ & $g_m^{\mpgtf}(t)$ & \textbf{Yes} & \textbf{Yes}
\\ \tthline
\end{tabular}
}
\label{tab:methods}
\end{table}

\begin{table}
\centering
\caption{Hyperparameters of masking modules used in experiments}
{
\begin{tabular}{c|c|c} \tthline
  Symbol & Description & Value \\ \thline
  $N$ & \# of channels of latent representation & $440$ \\
  \multirow{2}{*}{$B$} & \# of channels in bottleneck and & \multirow{2}{*}{$160$} \\ 
  & residual paths' $1\times1$ convolution blocks & \\ 
  \multirow{2}{*}{$Sc$} & \# of channels in skip-connection paths' & \multirow{2}{*}{$160$} \\
  & $1\times1$ convolution blocks & \\ 
  $H$ & \# of channels in convolution blocks & $160$ \\ 
  $P$ & Kernel size in convolution blocks & $3$ \\ \tthline
  \end{tabular}
  \label{tab:config}
}
\end{table}

\begin{figure*}[ht]
\centering
\includegraphics[width=1.1\columnwidth,clip]{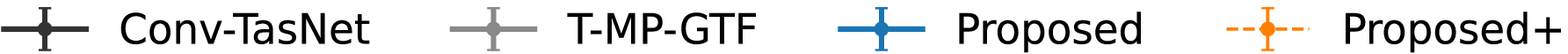}
\vspace{2mm}
\\
\includegraphics[width=2.0\columnwidth,clip]{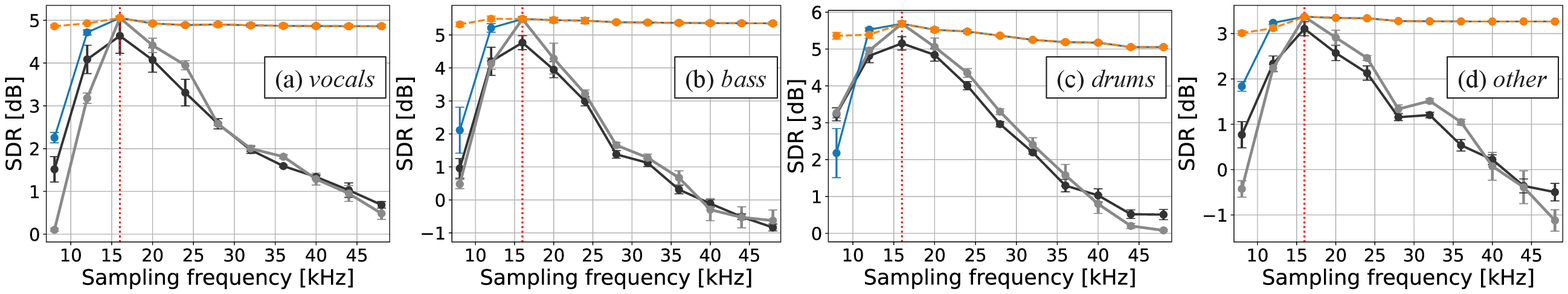}
 \caption{SDRs of Conv-TasNet and proposed models for test data at various sampling frequencies. These SDRs and error bars respectively denote averages and standard errors over results obtained with four random seeds. Red line shows trained sampling frequency.}
 \label{fig:SDR}
\end{figure*}

\mysubsection{Experimental Settings}
\label{ssec:experimentalsettings}
To evaluate the efficacy of the proposed method, we conducted music source separation on the MUSDB18-HQ dataset \cite{musdb18-hq}, which consists of $86$ training, $14$ validation, and $50$ test tracks.
Each track contains separate recordings of four musical instruments ({\em vocals}, {\em bass}, {\em drums}, and {\em other}), i.e., $C=4$.
The training and validation tracks were down-sampled to $16$ kHz, and we created the test data by down- and up-sampling the test tracks to several target sampling frequencies, $F_{s}^{\target}=8,12,\ldots,48$ kHz.
As an evaluation metric, we used the median signal-to-distortion ratios (SDRs) computed with the BSSEval v4 toolkit \cite{bsseval}.


We used the same data augmentation techniques as those of \cite{metatas}: random cropping of the $8$ s training audio segments, random amplification within $[0.75,1.25]$, random selection of the left or the right channel, and random intertrack shuffling of the instruments in half of the minibatch.
We also applied standardization (zero mean and unit variance) to the tracks.

We compared the proposed model (Proposed) and that using the aliasing reduction technique (Proposed+) with Conv-TasNet and its variant (T-MP-GTF) whose encoder and decoder instead use the trainable extension of MP-GTF as their weights of the convolution and transposed convolution layers, respectively.
T-MP-GTF was included to separately evaluate the trainable extension of the MP-GTF and the sampling frequency adaptation.
We applied all models to the audio signals of the unseen sampling frequencies without resampling them at the trained sampling frequency in order to examine the effects of the sampling frequency mismatch and the proposed sampling frequency adaptation.
Table~\ref{tab:methods} summarizes the features of these models.
For Proposed and Proposed+, we determined $L$ and $W$ to be $5.0$ and $2.5$ ms at the sampling frequency of $16$ kHz, respectively, as described in Section~\ref{ssec:adaptingsfv}, whereas we set $L=80$ and $W=40$ for the other models.
For all models, we set $X=6$ and $R=2$.
The hyperparameters of the masking modules are shown in Table~\ref{tab:config}, where the symbols correspond to those used in the literature of Conv-TasNet (see Table~1 in \cite{convtas}).


For $g_m^{\mpgtf}(t)$, we trained $f_m$ and $\phi_m$ for $m=1, \ldots , 220$ jointly with the entire network, and constrained these parameters for the other $m$s so that $f_{m+220}=f_m$ and $\phi_{m+220}=\phi_{m}+\pi$, as described in Section \ref{ssec:generativeprocess}.
We initialized $f_m$ and $\phi_m$ as in \cite{mpgtf};
let us denote $48$ frequencies distributed uniformly in the equivalent rectangular bandwidth (ERB) scale \cite{erb} from $50$ to $8000$ Hz by $f^{\mathrm{(center)}}_{i}$, where $i=1,\cdots,48$ is the center frequency index and $f_{i}<f_{i+1}$ for all $i$.
We initialized $f_m$ as $f_m=f^{\mathrm{(center)}}_{\lfloor m/K \rfloor+1}$ for $m=1,\ldots,140$ ($m=141,\ldots,220$), where $K=5$ ($K=4$, respectively).
The phase shifts $\phi_m$ of the filters with the same $f^{\mathrm{(center)}}_i$ were initialized to be uniformly distributed in $[0,\pi)$.
The other parameters were set as $a_m=1$, $p_m=2$.
As in \cite{mpgtf}, these filters were normalized so that they have the same $l^2$ norm.

For training, we used the RAdam optimizer \cite{radam} with a weight decay rate of $5.0\times10^{-4}$ and the Lookahead mechanism \cite{lookahead} with a synchronization period of $6$ and a slow weights step of $0.5$.
Gradient clipping with the maximum $L_2$-norm of $5$ was applied.
The learning rate scheduler presented in \cite{sgdr} was employed with an initial learning rate of $1.0\times10^{-3}$ and a restart period of $200\,000$ iterations.
We trained each model with a batch size of $12$ for $250$ epochs, using the negative scale-invariant source-to-noise ratio as the loss function, and selected the model with the lowest validation loss.
We applied the trained models to the left and right channels of the test tracks separately, and scaled the source estimates using instrument-wise factors to minimize the mean squared error between the input mixture and the sum of all instrument estimates, resulting from the scale invariance of the loss function \cite{metatas}.

\vspace{-0.2\baselineskip}
\mysubsection{Results}
\label{ssec:results}
\label{sssec:ablationstudy}
Fig.~\ref{fig:SDR} shows the separation performances for the test data with sampling frequencies from $8$ to $48$ kHz.
At the trained sampling frequency, $16$ kHz, the proposed models including T-MP-GTF achieved higher SDRs than Conv-TasNet for all instruments, showing the effectiveness of the proposed trainable extension of the MP-GTF.
Interestingly, the proposed models gave much lower standard errors than Conv-TasNet, which was not reported in \cite{mpgtf}.
This observation reveals that the use of the trainable MP-GTF makes Conv-TasNet robust to the initialization of the DNN parameters.

As the sampling frequency moved away from $16$ kHz, the SDRs of Conv-TasNet and T-MP-GTF greatly decreased and they failed to separate the sources.
By contrast, the proposed models with the sampling frequency adaptation, Proposed and Proposed+, provided similar SDRs for the $12$- to $48$-kHz-sampled data and outperformed the other models by a large margin, particularly at $20$ kHz and higher, even though they were trained only with the $16$-kHz-sampled data.
This result clearly shows that the proposed sampling frequency adaptation plays a crucial role in achieving the consistent performance.

\begin{figure}[t]
\centering
\includegraphics[width=0.95\columnwidth,clip]{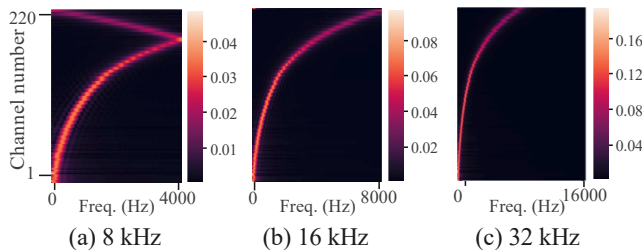}
\caption{Magnitudes of frequency responses of first $220$ filters of trained SFI convolution layer at sampling frequencies of $8, 16,$ and $32$ kHz.}
 \label{fig:encoder_visualization}
\end{figure}

Fig.~\ref{fig:encoder_visualization} shows the magnitudes of the frequency responses of the filters of the trained SFI convolution layer at sampling frequencies of $8,16,$ and $32$ kHz.
We can confirm that the filters of the $16$ and $32$ kHz sampling frequencies exhibited consistent frequency responses.
Importantly, the filters of the $32$ kHz sampling frequency blocked the frequency components higher than around $8$ kHz.
Nevertheless, the proposed model achieved a consistent performance at the sampling frequencies higher than $8$ kHz, which should be because the dominant frequency components of the music signals were distributed below $8$ kHz.

For the sampling frequency of $8$ kHz, aliasing occurred from $4$ kHz (see Fig.~\ref{fig:encoder_visualization}(a)).
This resulted in the performance degradation of Proposed (the proposed method without the aliasing reduction technique).
By contrast, Proposed+ showed a consistent performance at all sampling frequencies, demonstrating the effectiveness of the proposed aliasing reduction technique when the sampling frequency is reduced.
For {\em drums} and {\em other}, Proposed+ gave slightly lower SDRs than Proposed at the sampling frequency of $12$ kHz, which might be because the filters with center frequencies near the Nyquist frequency are helpful for the separation.
A further investigation of this observation remains as a future work.

\vspace{-0.1\baselineskip}
\mysection{Conclusion}
\vspace{-0.1\baselineskip}
We proposed an SFI convolution layer that can be adjusted to an arbitrary sampling frequency.
We focused on the fact that the weights of a convolution layer can be seen as a collection of digital FIR filters.
We explicitly defined the weights generation process of the convolution layer from the latent analog filters based on the impulse invariant method.
Since the analog filters do not depend on sampling frequency, the proposed layers can generate consistent weights for arbitrary sampling frequencies.
Furthermore, we built an SFI audio source separation model by incorporating the proposed layers into the encoder and decoder of Conv-TasNet.
We showed, through music source separation experiments, that even when trained with only the audio signals sampled at a specific sampling frequency, the proposed model consistently worked well with not only the trained sampling frequency but also unseen ones.
Since the proposed layer is a general component for audio processing, it would also be useful for various audio applications such as speech separation \cite{Sawada2019APSIPATSIP,8462507,7952154,convtas,mpgtf}.



\begin{thebibliography}{8}
\scriptsize
\setlength{\itemsep}{-0.1pt}

\bibitem{8462507}
Z.~{Wang}, J.~{Le Roux}, and J.~R. {Hershey},
\newblock ``{Alternative objective functions for deep clustering},''
\newblock in {\em Proceedings of IEEE International Conference on Acoustics,
  Speech and Signal Processing}, 2018, pp. 686--690.

\bibitem{7952154}
D.~{Yu}, M.~{Kolbæk}, Z.~{Tan}, and J.~{Jensen},
\newblock ``{Permutation invariant training of deep models for
  speaker-independent multi-talker speech separation},''
\newblock in {\em Proceedings of IEEE International Conference on Acoustics,
  Speech and Signal Processing}, 2017, pp. 241--245.

\bibitem{convtas}
Y.~{Luo} and N.~{Mesgarani},
\newblock ``{Conv-TasNet: Surpassing ideal time–frequency ,magnitude masking
  for speech separation},''
\newblock {\em IEEE/ACM Transactions on Audio, Speech, and Language
  Processing}, vol. 27, no. 8, pp. 1256--1266, 2019.

\bibitem{waveunet}
D.~{Stoller}, S.~{Ewert}, and S.~{Dixon},
\newblock ``{Wave-U-Net: A Multi-Scale neural network for end-to-end audio
  source separation},''
\newblock in {\em Proceedings of International Society for Music Information
  Retrieval Conference}, 2018, pp. 334--340.

\bibitem{defossez2019music}
A.~{D{\'e}fossez}, N.~{Usunier}, L.~{Bottou}, and F.~{Bach},
\newblock ``{Music source separation in the waveform domain},''
\newblock {\em arXiv preprint arXiv:1911.13254}, 2019.

\bibitem{9053934}
T.~{Nakamura} and H.~{Saruwatari},
\newblock ``{Time-domain audio source separation based on Wave-U-Net combined
  with discrete wavelet transform},''
\newblock in {\em Proceedings of IEEE International Conference on Acoustics,
  Speech and Signal Processing}, 2020, pp. 386--390.

\bibitem{mpgtf}
D.~{Ditter} and T.~{Gerkmann},
\newblock ``{A Multi-Phase Gammatone Filterbank for speech separation via
  TasNet},''
\newblock in {\em Proceedings of IEEE International Conference on Acoustics,
  Speech and Signal Processing}, 2020, pp. 36--40.

\bibitem{liu2020channelwise}
H.~{Liu}, L.~{Xie}, J.~{Wu}, and G.~{Yang},
\newblock ``{Channel-wise subband input for better voice and accompaniment
  separation on high resolution music},''
\newblock in {\em Proceedings of INTERSPEECH}, 2020.

\bibitem{Takeuchi2019ICASSP}
D.~Takeuchi, K.~Yatabe, Y.~Koizumi, Y.~Oikawa, and N.~Harada,
\newblock ``{Data-driven design of perfect reconstruction filterbank for
  DNN-based sound source enhancement},''
\newblock in {\em Proceedings of IEEE International Conference on Acoustics,
  Speech and Signal Processing}, 2019, pp. 596--600.

\bibitem{Krebs2016DownbeatTU}
F.~{Krebs}, S.~{B{\"o}ck}, M.~{Dorfer}, and G.~{Widmer},
\newblock ``{Downbeat tracking using beat synchronous features with recurrent
  neural networks},''
\newblock in {\em Proceedings of International Society for Music Information
  Retrieval Conference}, 2016, pp. 129--135.

\bibitem{7416164}
S.~{Sigtia}, E.~{Benetos}, and S.~{Dixon},
\newblock ``{An end-to-end neural network for polyphonic piano music
  transcription},''
\newblock {\em IEEE/ACM Transactions on Audio, Speech, and Language
  Processing}, vol. 24, no. 5, pp. 927--939, 2016.

\bibitem{9052987}
F.~{Pedersoli}, G.~{Tzanetakis}, and K.~M. {Yi},
\newblock ``{Improving music transcription by pre-stacking a U-Net},''
\newblock in {\em Proceedings of IEEE International Conference on Acoustics,
  Speech and Signal Processing}, 2020, pp. 506--510.

\bibitem{Yu2013ICLR}
D.~{Yu}, M.~{Seltzer}, J.~{Li}, J.~{Huang}, and F.~{Seide},
\newblock ``{Feature learning in deep neural networks - studies on speech
  recognition},''
\newblock in {\em Proceedings of International Conference on Learning
  Representations}, 2013.

\bibitem{8639610}
A.~{Narayanan}, A.~{Misra}, K.~C. {Sim}, G.~{Pundak}, A.~{Tripathi},
  M.~{Elfeky}, P.~{Haghani}, T.~{Strohman}, and M.~{Bacchiani},
\newblock ``{Toward domain-invariant speech recognition via large scale
  training},''
\newblock in {\em IEEE Spoken Language Technology Workshop}, 2018, pp.
  441--447.

\bibitem{8574905}
J.~{Gao}, J.~{Du}, and E.~{Chen},
\newblock ``{Mixed-bandwidth cross-channel speech recognition via joint
  optimization of DNN-based bandwidth expansion and acoustic modeling},''
\newblock {\em IEEE/ACM Transactions on Audio, Speech, and Language
  Processing}, vol. 27, no. 3, pp. 559--571, 2019.

\bibitem{Oppenheim2001DSP}
A.~V. {Oppenheim}, J.~R. {Buck}, and R.~W. {Schafer},
\newblock {\em Discrete-time signal processing},
\newblock Prentice Hall, Upper Saddle River, NJ, 2001.

\bibitem{metatas}
D.~{Samuel}, A.~{Ganeshan}, and J.~{Naradowsky},
\newblock ``{Meta-learning extractors for music source separation},''
\newblock in {\em Proceedings of IEEE International Conference on Acoustics,
  Speech and Signal Processing}, 2020, pp. 816--820.

\bibitem{Zeiler2014ECCV}
{M.~D.} Zeiler and R.~{Fergus},
\newblock ``{Visualizing and understanding convolutional networks},''
\newblock in {\em Proceedings of European Conference on Computer Vision}, 2014,
  pp. 818--833.

\bibitem{Gong2018Interspeech}
Y.~{Gong} and C.~{Poellabauer},
\newblock ``{Impact of aliasing on deep {CNN}-based end-to-end acoustic
  models},''
\newblock in {\em Proceedings of INTERSPEECH}, 2018, pp. 2698--2702.

\bibitem{Wu2018ECCV}
Y.~{Wu} and K.~{He},
\newblock ``{Group normalization},''
\newblock in {\em Proceedings of European Conference on Computer Vision}, 2018.

\bibitem{musdb18-hq}
Z.~{Rafii}, A.~{Liutkus}, F.-R. {Stöter}, S.~I. {Mimilakis}, and R.~{Bittner},
\newblock ``{{MUSDB18-HQ} - an uncompressed version of MUSDB18},'' 2019.

\bibitem{bsseval}
F.-R. {St{\"o}ter}, A.~{Liutkus}, and N.~{Ito},
\newblock ``The 2018 signal separation evaluation campaign,''
\newblock in {\em Proceedings of International Conference on Latent Variable
  Analysis and Signal Separation}, 2018, pp. 293--305.

\bibitem{erb}
V.~{Hohmann},
\newblock ``Frequency analysis and synthesis using a gammatone filterbank,''
\newblock {\em Acta Acustica united with Acustica}, vol. 88, no. 03, pp.
  433--442, 2002.

\bibitem{radam}
L.~{Liu}, H.~{Jiang}, P.~{He}, W.~{Chen}, X.~{Liu}, J.~{Gao}, and J.~{Han},
\newblock ``{On the variance of the adaptive learning rate and beyond},''
\newblock in {\em Proceedings of International Conference on Learning
  Representations}, 2020.

\bibitem{lookahead}
M.~{Zhang}, J.~{Lucas}, J.~{Ba}, and G.~{Hinton},
\newblock ``{Lookahead Optimizer}: k steps forward, 1 step back,''
\newblock in {\em Proceedings of Advances in Neural Information Processing
  Systems}, 2019, pp. 9597--9608.

\bibitem{sgdr}
I.~{Loshchilov} and F.~{Hutter},
\newblock ``{SGDR: Stochastic gradient descent with warm restarts},''
\newblock in {\em Proceedings of International Conference on Learning
  Representations}, 2017.
  
\bibitem{Sawada2019APSIPATSIP}
H.~{Sawada}, N.~{Ono}, H.~{Kameoka}, D.~{Kitamura}, H.~{Saruwatari},,
\newblock ``{A review of blind source separation methods: two converging routes to ILRMA originating from ICA and NMF},''
\newblock in {\em APSIPA Transactions on Signal and Information Processing}, vol.8, no. e12, 14 pages, 2019.

\end{thebibliography}
\end{document}